\documentclass[usenatbib,usegraphicx]{mn2e}
\usepackage{amstext}

\title[Satellite galaxies around Andromeda and $\Lambda$CDM simulations]
{A comparison of the distribution of satellite galaxies around Andromeda and the results of $\Lambda$CDM simulations}
\author[Bahl et al.]{H. Bahl$^{1}$\thanks{E-mail:
henning.bahl@tum.de}, H. Baumgardt$^{2}$\\
$^{1}$Physik-Department der Technischen Universit\"at M\"unchen, James-Franck-Str. 1, 85748 Garching, Germany\\
$^{2}$School of Mathematics and Physics, University of Queensland, St. Lucia, QLD 4072, Australia}

\begin{document}

\date{Accepted 2013 xx xx. Received 2013 xx xx; in original form 2013 xx xx}

\pagerange{\pageref{firstpage}--\pageref{lastpage}} \pubyear{201x}

\maketitle

\label{firstpage}

\begin{abstract}
 
\cite{I13} recently reported the existence of a vast thin plane of dwarf galaxies (VTPD) orbiting around Andromeda. We investigate whether such a configuration can be reproduced within the standard cosmological framework and search for similar planes of co-rotating satellite galaxies around Andromeda-like host haloes in data from the Millennium~II simulation combined with a semi-analytic galaxy formation model. We apply a baryonic mass cut of $2.8\times 10^4\text{M}_{\sun}$ for the satellite haloes and restrict the data
to a PAndAS like field. If we include the so-called orphan galaxies in our analysis, we find that planes with a rms lower than the VTPD are common in Millennium II. This is partially due to the strongly radially concentrated distribution of orphan galaxies. Excluding part of the orphan galaxies brings the radial distributions of Millennium II satellites into better agreement with the satellite distribution of Andromeda while still producing a significant fraction of planes with a lower rms than the VTPD.  We also find haloes in Millennium II with an equal or higher number of co-rotating satellites than the VTPD. This demonstrates that the VTPD is not in conflict with the standard cosmological framework, although a definite answer of this question might require higher resolution cosmological simulations that do not have to consider orphan galaxies.  Our results finally show that satellite planes in Millennium II are not stable structures, hence the VTPD might only be a statistical fluctuation of an underlying more spherical galaxy distribution.

\end{abstract}


\section{Introduction}

The distribution of the classical satellite galaxies around the Milky Way is highly anisotropic with most of them being located in a thin plane roughly perpendicular to the disc of the Milky Way \citep{K05}. \citet{M08} found that the clustering of the orbital poles indicates a rotational support of the plane of satellite galaxies. \citet{L05}, \citet{Z05}, \citet{L09}, \citet{L11} and \citet{W13} used large-scale cosmological N-body simulations and tried to find similar anisotropic distributions of satellite galaxies. \citet{L05} noted that in all six of their simulations of dark matter haloes the 11 brightest satellites are distributed along thin, disc-like structures and that this distribution reflects the preferential infall of satellites along the spines of a few filaments of the cosmic web. \citet{Z05} found that if selecting the subhaloes expected to be luminous, the distribution of the subhaloes around the three simulated Milky Way-sized dark matter haloes is consistent with the observed Milky Way satellites due to the preferential accretion of satellites along filaments and the evolution of satellite orbits within prolate, triaxial potentials. \citet{L09} simulated 436 haloes and found that the angular momenta of at least three of the eleven brightest satellites point approximately towards the short axis of the satellite distribution in a significant fraction of the haloes. This result is supported by \citet{L11}, who found that quasi-planar distributions of coherently rotating satellites, such as those inferred in the Milky Way and other galaxies, arise naturally in simulations of a $\Lambda$CDM universe. \citet{W13} examined 1686 Milky Way-like haloes found in the Millennium~II simulation \citep{B09} and six haloes simulated by the Aquarius project \citep{S08}. They noted that 5-10 per cent of the satellite distributions are as flat as the Milky Way's and that this can be best understood as the consequence of preferential accretion along filaments of the cosmic web and the accretion of a single rich group of satellites. Nevertheless,
 \citet{P12} argued that CDM simulations do not reproduce the clustering of the orbital poles of the Milky Way satellites. 

The distribution of satellite galaxies around Andromeda is also known to be anisotropic \citep{Ko06, M06}. \citet{I13} and \citet{C13} recently noted the existence of a vast thin plane of dwarf galaxies (VTPD) orbiting around the Andromeda galaxy. They analysed 27 satellites examined by the Pan-Andromeda Archaeological Survey (PAndAS, \citealt{Mc09}) and found that 15 out of the 27 satellites are located in a thin plane with root mean square thickness of $12.6\pm 0.6$ kpc. In addition, 13 of these 15 satellites are supposed to share the same sense of rotation around Andromeda \citep{I13}. \citet{I13} found that the existence of this thin plane is highly significant. By assigning a random angular orientation inside the PAndAS area to each satellite galaxy, \citet{I13} found that only in 0.13 per cent of all cases a plane as thin as the VTPD is found. \citet{I13} discussed accretion of satellite galaxies and the in-situ formation as two scenarios to create such a thin plane. They argue that both do not seem to be able to explain the existence of the VTPD.      

A possible clustering of the orbital poles of Andromeda's satellites cannot be examined since the proper motions of the Andromeda satellites are not yet known. Nevertheless a large number of co-rotating satellites located in a thin plane is a remarkable similarity between the Andromeda galaxy and the Milky Way and might point to a similar formation history \citep{H13}. This idea is backed by the observations of \citet{I13} that the identified plane is approximately aligned with the pole of the Milky Way's disk and is co-planar to the position vector between the Milky Way and Andromeda. On the other hand it is of course possible that Andromeda and the Milky Way formed independently and that the alignment is just random. Hence the question remains, whether standard galaxy formation models within the current cosmological framework are sufficient to understand the formation of the Andromeda and Milky Way satellite distributions.

We try to answer this question here for Andromeda. Our approach is to search for similar structures in the Millennium~II simulation \citep{B09}, and to examine their formation and evolution. \citet{G13} scaled the Millennium~II data to match the cosmological parameters derived by WMAP7 \citep{J11} and ran a semi-analytic galaxy formation model using the formalism of \citet{G11} over the re-scaled data.
The main difference between the approaches by \citet{L05} and \citet{Z05} and the approach used here is the number of examined host haloes. The Millennium~II simulation provides a large sample of Andromeda like host haloes. This high number makes it possible to quantify the commonness of such structures more effectively. In addition, we investigate the formation history of some haloes featuring a thin plane. Due to the rescaling, the scaled Millennium~II simulations does not stop at present time. Therefore we are also able to examine the further evolution of thin planes. The approach presented here can be used to test whether galaxy formation within the $\Lambda$CDM paradigm is sufficient to explain the formation of such structures. Our paper is organised as follows: In section 2 we discuss our method for selecting Andromeda like satallite distributions from the Millenium II simulation. In section 3, we present the results that we obtained using this method. We end with a conclusion and discussion of the results in section 4.

\section{Methods}

The scaled version of the Millennium~II simulation was created by \citet{G13} using the algorithm of \citet{A10}. This algorithm involves first a reassignment of length, mass and velocity units, followed by a relabelling of the time axis and finally a rescaling of the amplitudes of individual large-scale fluctuation modes. The re-scaled simulation has a box size of 104.3 Mpc/h containing $2160^3$ particles with masses of $8.5024\times 10^6 \text{M}_{\sun}$/h. It uses the following cosmological parameters: matter density $\Omega_m = 0.272$, baryon density $\Omega_b = 0.045$, dark energy density $\Omega_\Lambda = 0.728$, Hubble parameter $\text{h}=0.704$, scalar spectral index $n_s = 0.961$ and a fluctuation amplitude at 8 h$^{-1}$Mpc of $\sigma_8 = 0.807$.  We use the data derived by a semi-analytic galaxy formation model run on the scaled Millennium~II data here. \citet{G13} used the semi-analytic galaxy formation model developed by \citet{G11} which considers the influence of gas infall (both cold and hot, primordial and recycled), shock heating, cooling, star formation, stellar evolution, supernova feedback, black hole growth, AGN feedback, metal enrichment, mergers and tidal and ram-pressure stripping. 

The resolution of the Millennium II simulation is high enough to resolve sub-haloes down to masses of $2\times 10^8 \text{M}_{\sun}$ \citep{B09}. Below this limit, individual satellite haloes contain less than 20 particles and are not resolved any more. Guo et al. (2011) assume that these galaxies are still bound and determine the position of these so-called 'orphan galaxies' by tracking the most bound particle before the disruption of the subhalo. The validity of this model can be questioned, in particular it is likely that at least some of the orphan galaxies are tidally disrupted. The positions and velocities of the orphan galaxies might also not be accurate. Simulations with higher resolution would be necessary to solve this problem as indicated by \citet{G11}, however there are indications that the approach by Guo et al. still gives reliable results down to lower halo masses. For example \citet{W13} compared the radial distributions of satellite systems in the Millennium II simulation with those in the Aquarius simulation \citep{S08} and the radial distribution of the classical Milky Way satellites. They found that the radial distribution predicted by the Millennium II simulation is in good agreement with the higher resolution Aquarius simulations as well as with the observations as long as orphan galaxies are included in the analysis. In addition, \citet{G11} found that the stellar mass and luminosity functions predicted by their semi-analytic galaxy formation model match observations well if orphan galaxies are included. In case of the Milky Way for example they found that the total number of galaxies brighter than M$_\text{V}\sim -5$ agrees well with the number of Milky Way satellites brighter than this magnitude limit as predicted by \citet{K08}.

In a first step, we query possible host haloes from the Millennium~II database. The observed virial mass of Andromeda is $\sim 1.4\times 10^{12} \text{M}_{\sun}$ \citep{W10}, therefore we select all haloes from Millennium~II with a virial mass between $1.1 \times 10^{12} \text{M}_{\sun}$ and $1.7 \times 10^{12} \text{M}_{\sun}$. We exclude haloes whose mass weighted age according to the semi-analytic calculations of \citet{G13} is higher than 10 Gyr, since the average stellar age of stars in the disc of Andromeda is supposed to be less than 10 Gyr \citep{B06}. We reject any host halo which has a satellite galaxy with a baryonic mass higher than $7\times 10^{10} \text{M}_{\sun}$ within a spatial distance of 500~kpc, because such galaxies are not found near Andromeda and might disturb the distribution of dwarf galaxies around the host halo. 1825 haloes fulfil these requirements and are regarded as host haloes. For our analysis, we retrieve all relevant parameters (virial and baryonic masses, positions, velocities) of dwarf galaxies around each host halo. These dwarf galaxies are used to search for planes of satellite galaxies.

Faint dwarf galaxies are hard to detect. Dwarf galaxies in the PAndAS field have baryonic masses down to about $2.9\times 10^4 \text{M}_{\sun}$ \citep{M12}. To take care of this mass limit in our analysis we remove all dwarf galaxies with a baryonic mass less than $2.8\times 10^4 \text{M}_{\sun}$ from the data.

\citet{I13} only considered satellites within the PAndAS area \citep{Mc09}. Due to the luminous disk of Andromeda they discarded satellites within a projected distance of 32 kpc as well. We approximate the PAndAS area as a sphere with a radius of 128 kpc. Since there is no particular line of sight from which satellite galaxies are seen in the Millennium~II simulation, we choose one randomly. We assign an approximated PAndAS area to each line of sight and exclude all satellites outside this area. We also discard all satellites within a projected distance of 32 kpc to take obscuration due to a luminous disk into account. For haloes where too many satellites are excluded, we choose a new line of sight three times until a sufficient number of satellite galaxies is found. If after three tries we still cannot find enough satellite galaxies, we discard the whole halo.

We end up with a list of satellite galaxies with baryonic masses greater than $2.8\times 10^4 \text{M}_{\sun}$ which are located in a PAndAS like area for each host halo having an average virial mass of $10^{12} \text{M}_{\sun}$. We find that haloes in the Millennium~II simulation have on average 48 subhaloes fulfilling the requirements described above in a PAndAS like field. In order to compare the distribution
of Millennium~II halos  with the Andromeda satellite system, we follow two different approaches. The first is based on the assumption that the PAndAS is not complete. This is not an unreasonable assumption, as indicated by the large number of new satellites that were discovered by PAndAS itself. Most of the remaining satellites (if any) are probably close to the faint end of PAndAS. In order to perform a comparable analysis to that of \citet{I13}, we select the 27 satellites with the highest baryonic mass for each Millennium~II halo to end up with the same absolute number of satellites as \citet{I13}. Of these 27 satellites with the highest baryonic mass on average 67 per cent are orphan galaxies. The second approach is based on the assumption that PAndAS is (nearly) complete and that Andromeda is a halo with a lower than average number of satellite haloes. In this case, we select only those Millennium~II haloes with exactly 27 subhaloes fulfilling the requirements described above ($\sim 6$ per cent of all haloes). In this case on average 70 per cent of the satellites are orphan galaxies.

In order to see how this high fraction of orphan galaxies affect the results, we perform an additional analysis in which all orphan galaxies are excluded applying the first approach. In this case the average number of satellites is 12 and only 6 per cent of all haloes have 27 or more satellites and can be compared with Andromeda.

\cite{I13} searched for the plane which minimizes the root mean square distance $\text{r}_{\text{per}}$ of the satellite galaxies perpendicular to this plane. For the calculation of $\text{r}_{\text{per}}$ they only considered those 15 satellite galaxies which are closest to the plane. They mentioned that the number 15 was chosen only because the plane thickness increases significantly if a larger number of satellites is chosen.

When measuring the plane thickness, \citet{I13} took distance errors to individual galaxies into account. They calculated a $\text{r}_{\text{per}}$ value of $12.6 \pm 0.6$~kpc  for the Andromeda satellite distribution and we use this value as the $\text{r}_{\text{per}}$ value of Andromeda in this paper. We also use the 15 nearest satellites for the calculation of the best fitting plane for the Millennium~II haloes to ensure comparability. 

We characterize the satellite distribution around each halo by two different numbers. First, we calculate the root mean square distance $\text{r}_{\text{per}}$ of the satellites to the best fitting plane using the method of \citet{I13} descriped above. Second, we calculate the root mean square distance $\text{r}_{\text{par}}$ of the satellites projected in the plane to the center of the host halo. For the calculation of both $\text{r}_{\text{per}}$ and $\text{r}_{\text{par}}$, we use only the 15 satellites which are closest to the best fitting plane. It is important to compare the radial concentration of the satellite distributions of the Millennium~II haloes and the satellite distribution around Andromeda since it is of course more likely to find planes with low values of $\text{r}_{\text{per}}$ in more radially concentrated galaxy distributions. We use $\text{r}_{\text{par}}$ to measure the radial concentration. The $\text{r}_{\text{par}}$ value of the Andromeda satellite distribution turns out to be 167.5 kpc.

We furthermore calculate the sense of rotation of each satellite galaxy belonging to the best fitting plane. We calculate the sense of rotation of each satellite galaxy relative to the best fitting plane and determine the number of co-rotating and counter-rotating galaxies. The number of co-rotating galaxies is then given by the maximum of the two numbers. 

All the methods described above only use data from the snapshot closest to the present age of the universe. To learn about the time evolution of thin planes, we consider all haloes with an $\text{r}_{\text{per}}$ value lower than 14 kpc and with 13 or more co-rotating satellites and trace back the 15 satellites closest to the corresponding best fitting plane of these haloes to $z\sim 0.5$ corresponding to a time $\sim 5.2$ Gyr before present time ($T-T_H\sim -5.2$ Gyr) by querying their first progenitors and calculating the value of $\text{r}_{\text{per}}$ at every snapshot. The first progenitor is the most massive progenitor of the satellite in the majority of cases. We refer the readers to \citet{D07} for a detailed description of the algorithm choosing the first progenitor. If a first progenitor does not exist for a satellite, we reduce the number of satellites considered in the calculation of $\text{r}_{\text{per}}$ to the remaining number of satellites. A reduction of the number of considered satellites alleviates finding a good fitting plane to the remaining ones. In consequence lower $\text{r}_{\text{per}}$ values are more likely. 

In addition, we follow the further evolution of the 15 satellites by pursuing their descendants until $z\sim -0.29$ and calculating the value of $\text{r}_{\text{per}}$ at every snapshot. This is possible because the scaled version of the Millennium~II simulation does not stop at $z=0$ but at $z\sim -0.29$ corresponding to $T-T_H\sim 5.0$ Gyr. Due to mergers and the fact that not all haloes can be resolved in all snapshots, the number of satellites can also decrease if following the further evolution. In this case we reduce the number of satellites considered in the calculation of $\text{r}_{\text{per}}$ to the remaining number of satellites as in the case of the first progenitors.

\begin{figure}
\begin{center}
\includegraphics[width=95mm]{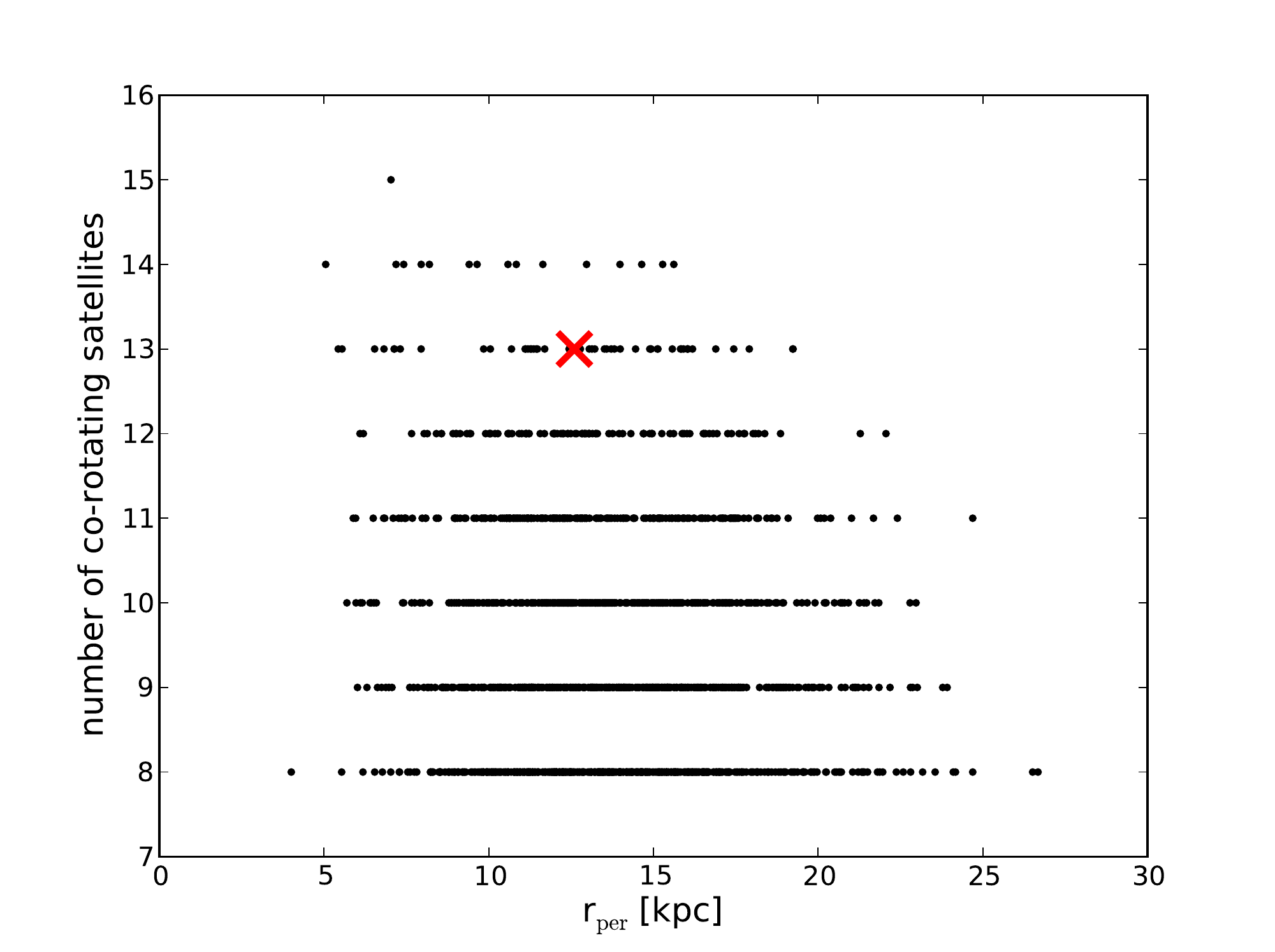}
\end{center}
\caption{The root mean square distance $\text{r}_{\text{per}}$ of the satellites to the best fitting plane plotted against the number of co-rotating satellites, if orphan galaxies are included. The black dots show individual haloes and their satellite systems from the Millennium~II simulation, the red cross marks the position of the satellite system of Andromeda. 40 per cent of the Millennium~II haloes have an $\text{r}_{\text{per}}$ value equal to or smaller than the $\text{r}_{\text{per}}$ of Andromeda.} 
\label{rper_rot}
\end{figure}

\begin{figure}
\begin{center}
\includegraphics[width=95mm]{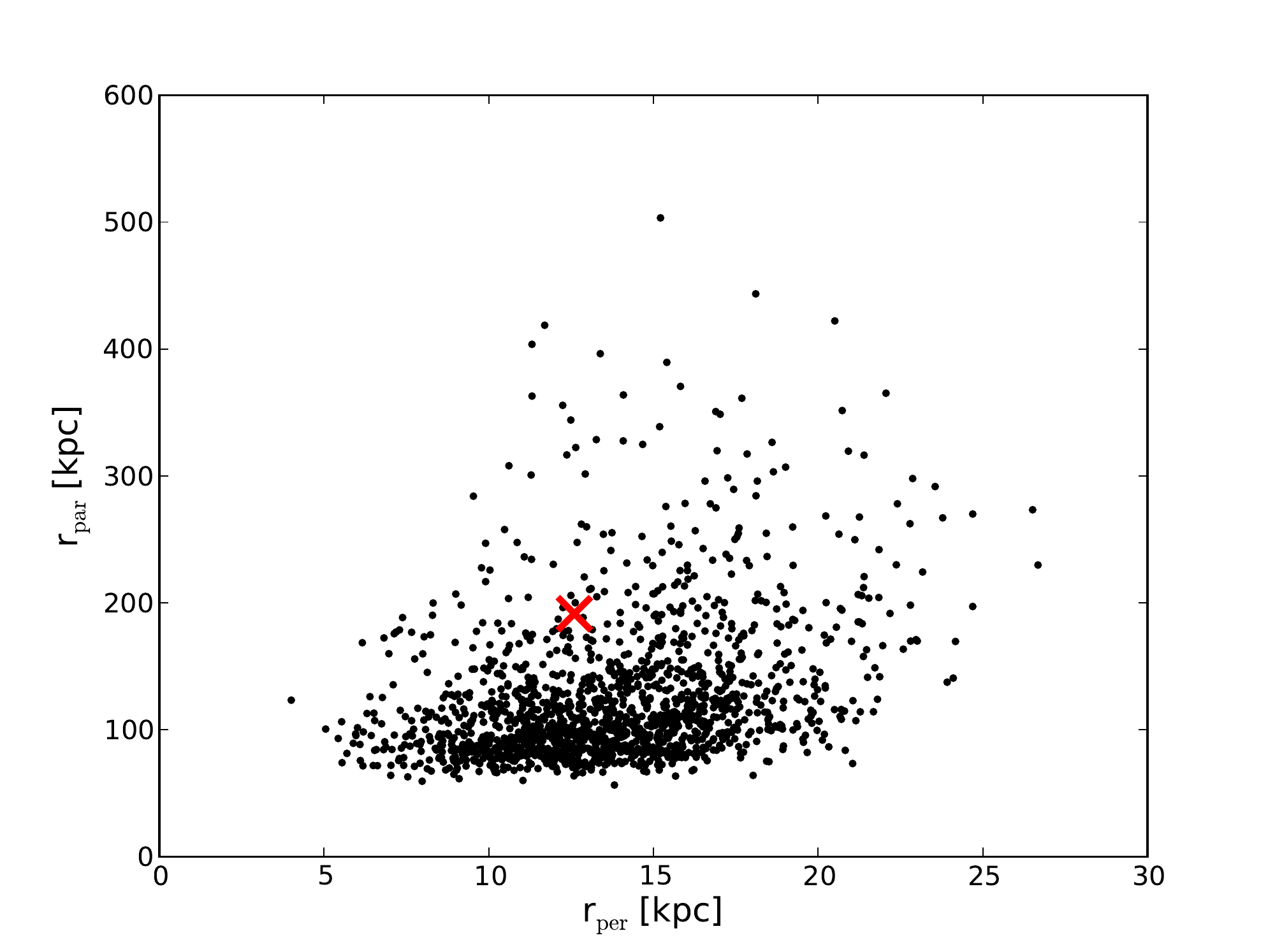}
\end{center}
\caption{The root mean square distance $\text{r}_{\text{per}}$ of the satellites to the best fitting plane plotted against the root mean square distance of the satellites projected in the best fitting plane to the center of the host halo ($\text{r}_{\text{par}}$), if orphan galaxies are included. The black dots show individual haloes and their satellite systems from the Millennium~II simulation, the red cross marks the position of the satellite system of Andromeda. 2 per cent of the Millennium~II haloes have a smaller $\text{r}_{\text{per}}$ and a higher $\text{r}_{\text{par}}$ than Andromeda.} 
\label{rper_rpar}
\end{figure}

\begin{figure}
\begin{center}
\includegraphics[width=95mm]{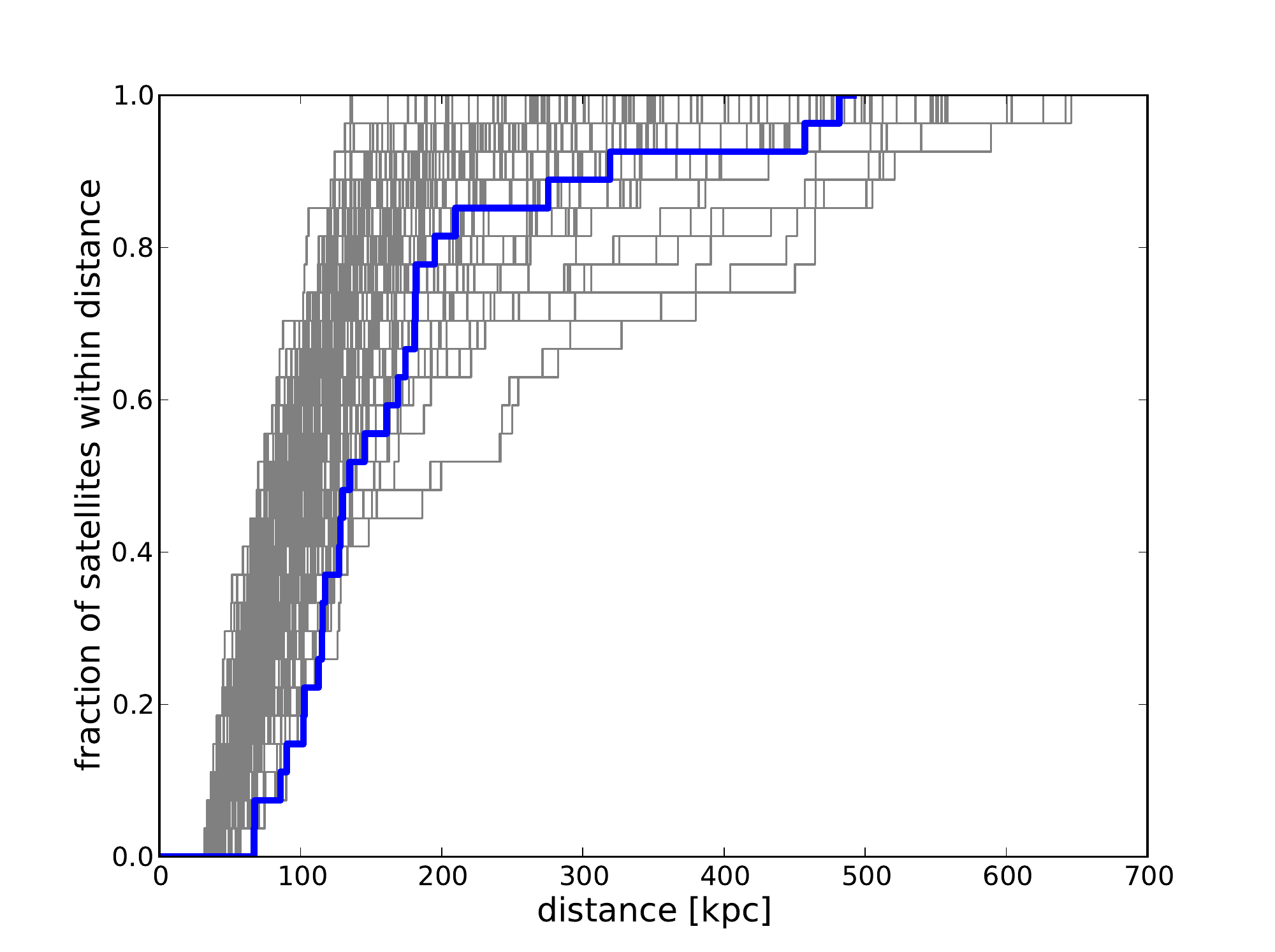}
\end{center}
\caption{Cumulative plot of the number of satellites within a certain distance from the center of the host halo, if orphan galaxies are included. The grey thin lines mark the profiles for 100 randomly selected Millennium~II haloes. The blue thick line is the profile of the satellite distribution around Andromeda.} 
\label{profiles}
\end{figure}

\begin{figure}
\begin{center}
\includegraphics[width=95mm]{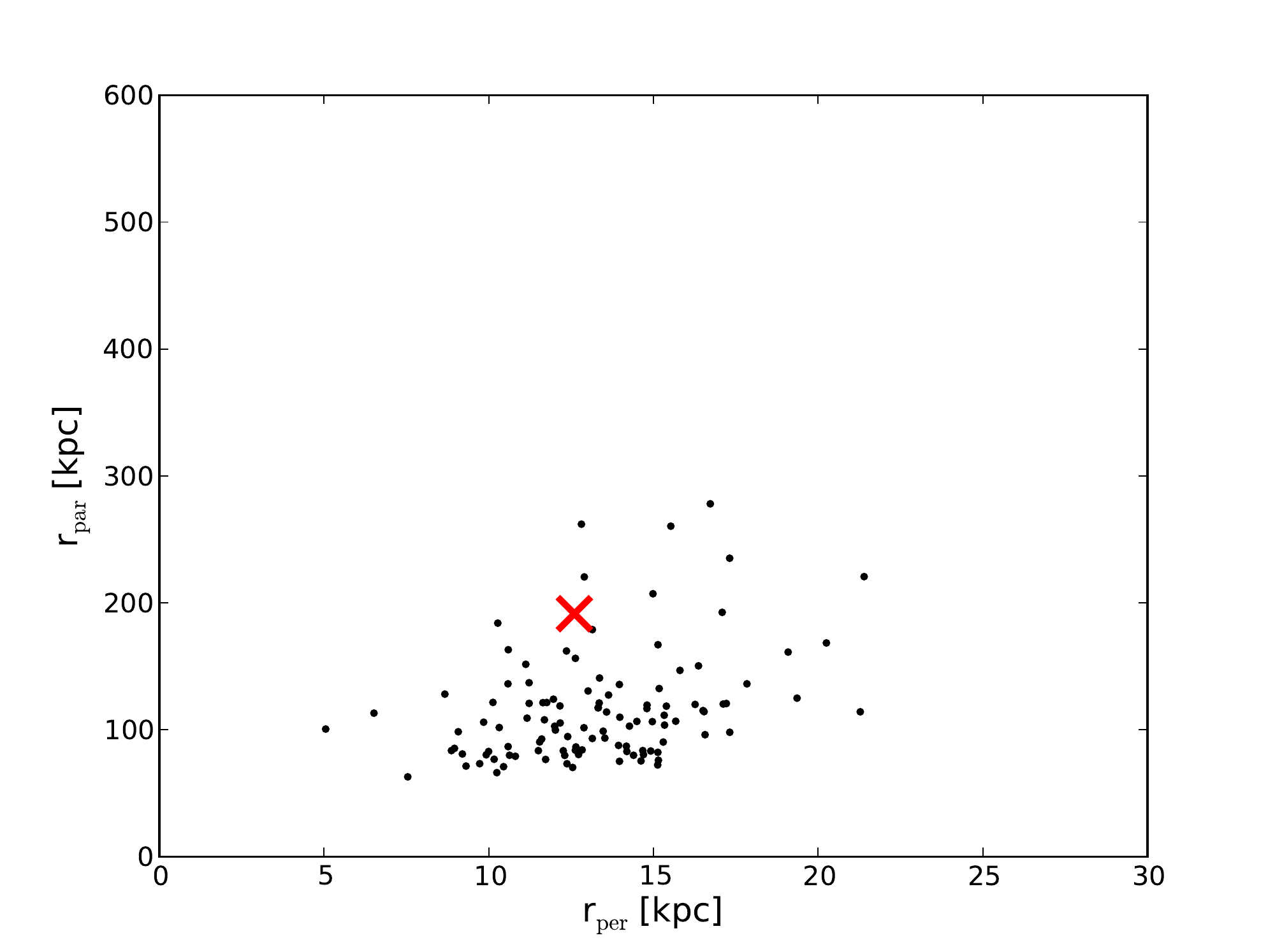}
\end{center}
\caption{Same as Fig. \ref{rper_rpar} but selecting only haloes with exactly 27 satellites. None of the Millennium~II haloes has a smaller $\text{r}_{\text{per}}$ and a higher $\text{r}_{\text{par}}$ than Andromeda.} 
\label{27rper_rpar}
\end{figure}

\section{Results}

We divide this section into three subsections: First we present the results obtained if all orphan galaxies are included. In sec. 3.2 we present the corresponding results if part or all of the orphan galaxies are excluded. Finally, in sec. 3.3 we present additional results about the number of corotating satellites, the evolution and further development of thin planes and the robustness against the adopted cosmology.

\subsection{Including orphan galaxies}

The results of our analysis if orphan galaxies are included are depicted in Figs. \ref{rper_rot} to \ref{profiles}. In total 1511 haloes remain after imposing the various selection criteria. Of these 1511 haloes, 40 per cent have an $\text{r}_{\text{per}}$ value equal to or smaller than the $\text{r}_{\text{per}}$ value of Andromeda. 2 per cent of all haloes have simultaneously an $\text{r}_{\text{per}}$ value equal to or smaller than the $\text{r}_{\text{per}}$ value of Andromeda and a number of co-rotating satellites equal to or higher than the number of co-rotating satellites for the VTPD (see Fig. \ref{rper_rot}). These values are in stark contrast to a spherical distribution (inside the PAndAS area), for which \citet{I13} found that only 0.13 per cent of all distributions have a smaller $\text{r}_{\text{per}}$ value than Andromeda.

A valid objection to the good fit obtained in Fig. \ref{rper_rot} could be that the Millennium~II haloes are biased by a more radially concentrated distribution, since finding a halo with a low $\text{r}_{\text{per}}$ value is more likely for a more radially concentrated distribution (lower $\text{r}_{\text{par}}$ value). In order to test if this is the case, we plot in Fig. \ref{rper_rpar} the distribution of the root mean square distance perpendicular to the best-fitting plane ($\text{r}_{\text{per}}$) vs. the root mean square distance to the host halo in the best fitting plane ($\text{r}_{\text{par}}$). Fig. \ref{rper_rpar} shows that 10 per cent of all Millennium~II haloes have a larger $\text{r}_{\text{par}}$ than Andromeda and only 2 per cent of the haloes have simultaneously a smaller $\text{r}_{\text{per}}$ and a higher $\text{r}_{\text{par}}$ than Andromeda. This indicates that the good fit obtained in Fig. \ref{rper_rot} is at least partially due to the more radially concentrated satellite distribution of the Millennium~II haloes.

This result is confirmed by Fig. \ref{profiles}, which shows the radial distribution of satellites in 100 randomly selected haloes compared to the one of Andromeda. For nearly all haloes the satellites tend to be more radially concentrated than the one of Andromeda (a \mbox{K-S} test yields on average a probability of $\sim 16$ per cent that a profile from Millennium~II and the profile of Andromeda's satellite distribution are drawn from the same distribution) over the whole distance range. This could point to the fact that some of the orphan galaxies that we include in our analysis have actually been tidally disrupted, since orphan galaxies tend to be more radially concentrated than normal galaxies \citep{G11}. Lowering the number of orphan galaxies might help to decrease the average number of satellites as well as to improve the agreement between the radial distributions.

The previous results were obtained for haloes in which we have selected the 27 brightest galaxies. However, the results do not significantly change if we restrict ourselves to haloes that have exactly 27 satellite halos. We find in total 111 haloes fulfilling this requirement. In 41 per cent of these haloes the root mean square distance perpendicular to the best fitting plane ($\text{r}_{\text{per}}$) calculated by again considering only the best-fitting 15 satellites is smaller than 12.6 kpc. 2 per cent of all haloes have an $\text{r}_{\text{per}}$ value equal to or smaller than the $\text{r}_{\text{per}}$ value of Andromeda and a number of co-rotating satellites equal to or higher than the number of co-rotating satellites for the VTPD.

7 per cent of the haloes have a higher $\text{r}_{\text{par}}$ value than Andromeda. None of the haloes has a smaller $\text{r}_{\text{par}}$ and a higher $\text{r}_{\text{per}}$ value than Andromeda (see Fig. \ref{27rper_rpar}), which might however be due to the relatively small number of haloes. The closeness of at least one halo to Andromeda in Fig. \ref{27rper_rpar} indicates that if the sample of haloes was larger, most likely a non-vanishing fraction of haloes would be found that have a smaller $\text{r}_{\text{per}}$ value and a larger $\text{r}_{\text{par}}$ value than Andromeda. The radial distributions of the haloes only marginally match the radial distribution of Andromeda, a K-S test yields on average a probability of only 6 per cent that a profile from Millennium~II and the profile of Andromeda's satellite distribution are drawn from the same distribution.

To summarize, the exact way how 27 galaxies are selected does not seem to significantly influence the outcome of our analysis. If we include
orphan galaxies, we find a significant number of halos that have a smaller $r_{per}$ value than the VTPD, however the satellite distribution with
orphan galaxies included is also more radially concentrated than Andromeda's satellite distribution. Such a discrepancy in the radial distributions 
of satellite galaxies between Andromeda and cosmological simulations was also noted by \citet{R11}, who found that the Andromeda satellites within the
PAndAS field follow a radial distribution that falls off with distance as $1/r$, while cosmological simulations predict a steeper decrease.

\subsection{Excluding orphan galaxies}

\begin{figure}
\begin{center}
\includegraphics[width=95mm]{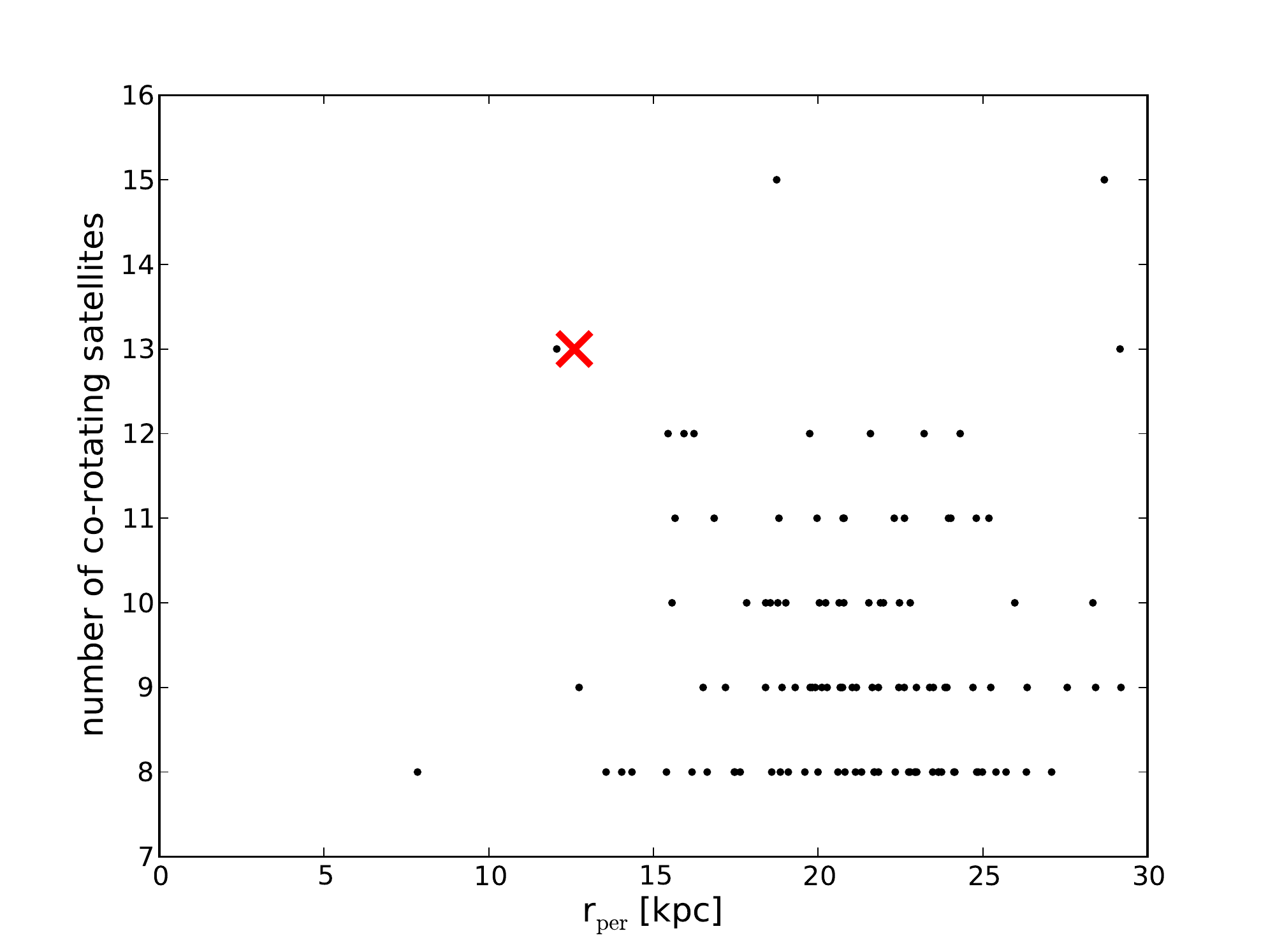}
\end{center}
\caption{Same as Fig. \ref{rper_rot} but excluding all orphan galaxies. 2 per cent of the Millennium~II haloes have an $\text{r}_{\text{per}}$ value equal to or smaller than the $\text{r}_{\text{per}}$ of Andromeda.} 
\label{rper_rot_withoutorphans}
\end{figure}

If we do not consider orphan galaxies in our analysis only 112 haloes fulfil the requirement that at least 27 satellites are located within a PAndAS like field. 
Fig. \ref{rper_rot_withoutorphans} depicts the root mean square distance perpendicular to the best fitting plane ($\text{r}_{\text{per}}$) vs. the number of corotating satellites
if only the 27 most luminous satellites are considered in these halos. Compared to Fig. \ref{rper_rot}, the distribution is now shifted towards higher $\text{r}_{\text{per}}$ values, the average $\text{r}_{\text{per}}$ has moved from 13.8 kpc with orphans to 21.12 per cent without orphans. 2 per cent of the haloes without orphans have an $\text{r}_{\text{per}}$ value lower than the one of Andromeda. 1 per cent simultaneously have 13 or more corotating satellites (see Fig. \ref{rper_rot_withoutorphans}). Excluding orphan galaxies, we therefore
find considerably less thin planes that are comparable to the VTPD, than if orphan galaxies are included.

\begin{figure}
\begin{center}
\includegraphics[width=95mm]{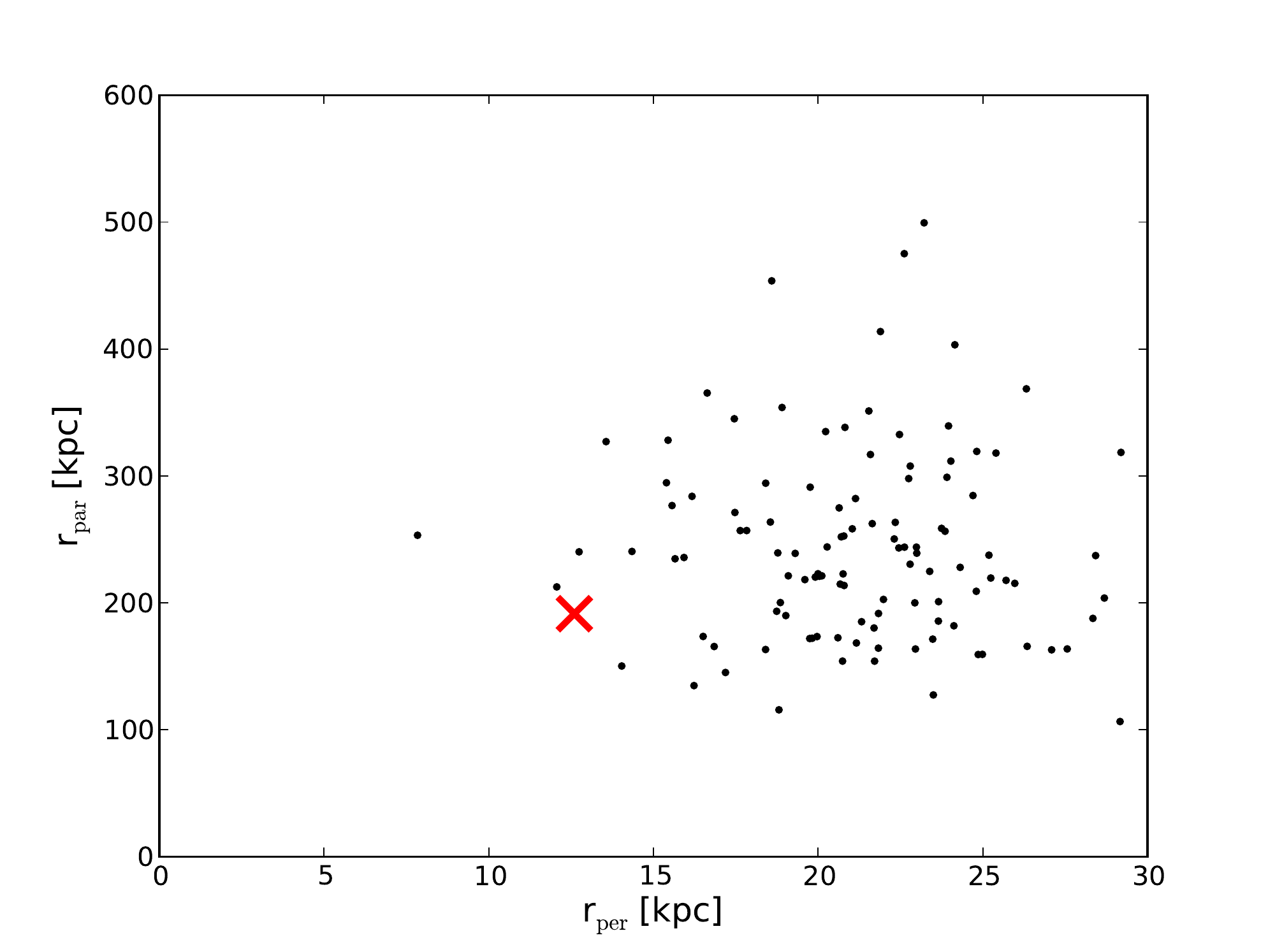}
\end{center}
\caption{Same as Fig. \ref{rper_rpar} but excluding all orphan galaxies. 73 per cent of the Millennium~II haloes have a higher $\text{r}_{\text{par}}$ than Andromeda. 2 per cent of the haloes simultaneously have a lower $\text{r}_{\text{per}}$ than Andromeda.} 
\label{rper_rpar_withoutorphans}
\end{figure}

\begin{figure}
\begin{center}
\includegraphics[width=95mm]{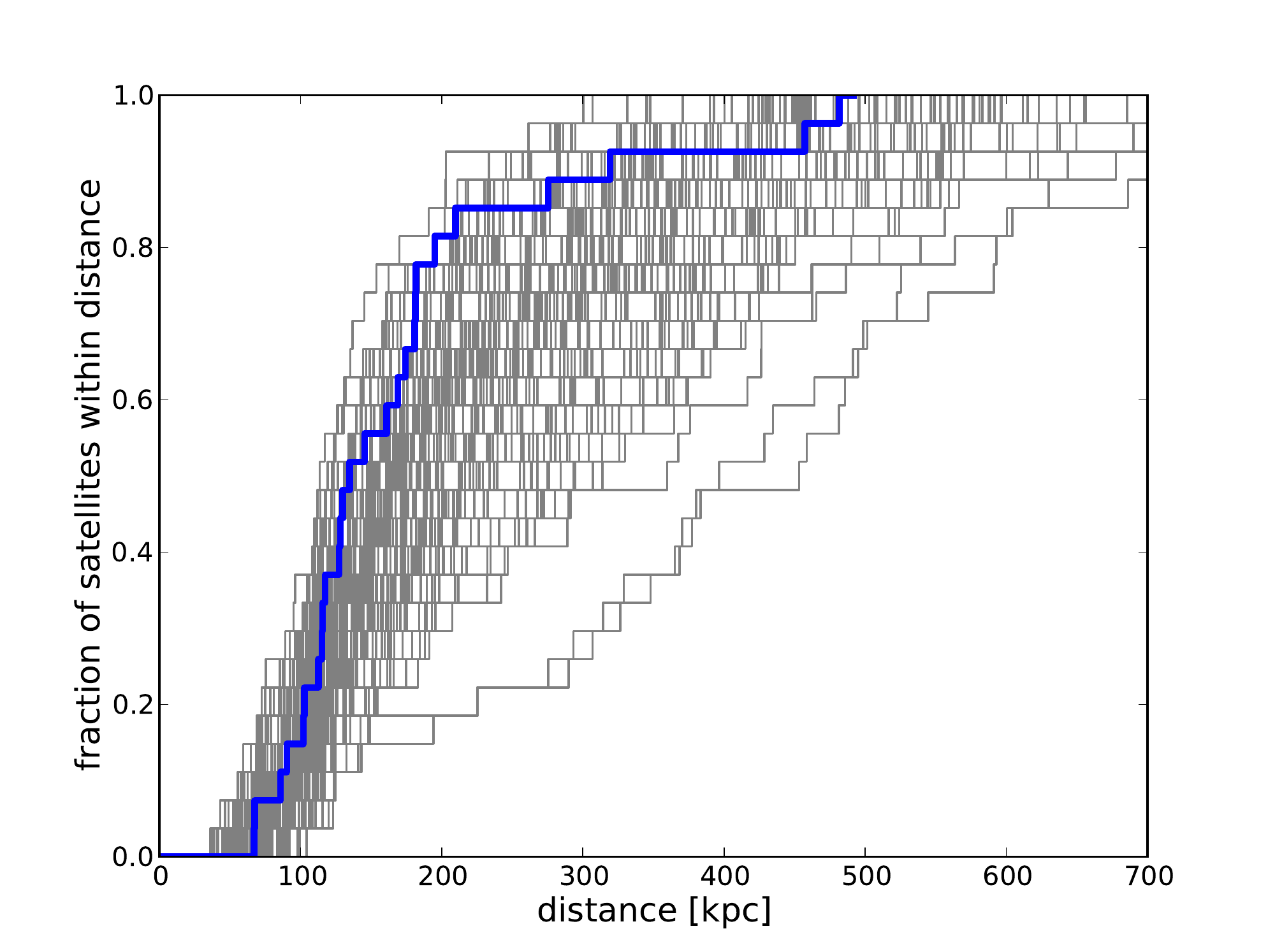}
\end{center}
\caption{Same as Fig. \ref{profiles} but excluding all orphan galaxies.} 
\label{profiles_withoutorphans}
\end{figure}

However without orphan galaxies the radial distributions agree much better. Fig. \ref{rper_rpar_withoutorphans} shows that 73 per cent of the haloes have an $\text{r}_{\text{par}}$ value higher than the one of Andromeda. 2 per cent of the haloes simultaneously have a lower $\text{r}_{\text{per}}$ value than Andromeda. The radial distributions are in good agreement with the one of Andromeda (see Fig. \ref{profiles_withoutorphans}). A \mbox{K-S} test yields on average a probability of $\sim 30$ per cent that a profile from Millennium~II and the profile of Andromeda's satellite distribution are drawn from the same distribution. The distributions match especially well below 150 kpc. For higher radii the Andromeda satellites are on average more radially concentrated than the Millennium~II satellites.

\begin{figure}
\begin{center}
\includegraphics[width=95mm]{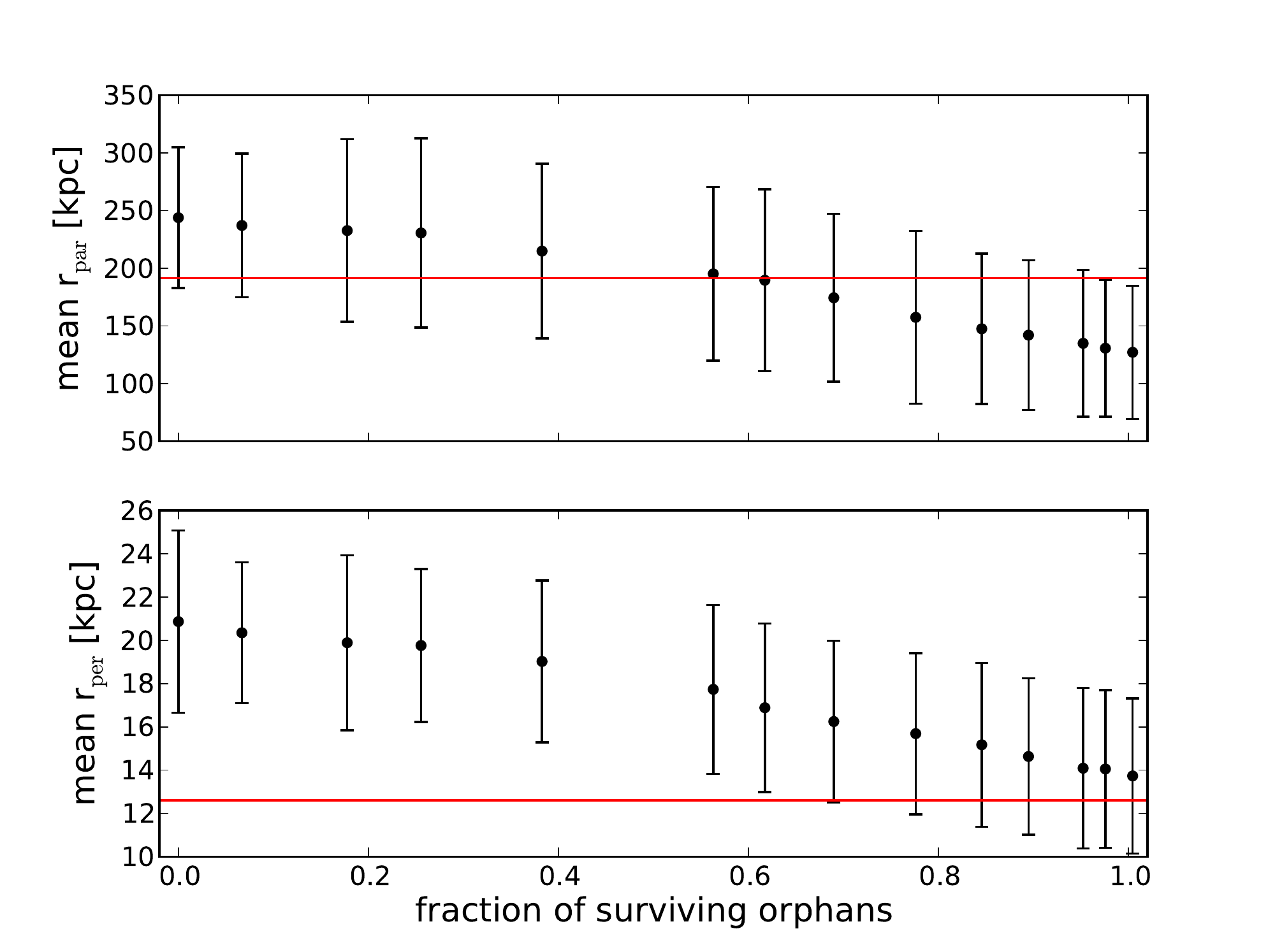}
\end{center}
\caption{Dependence of the mean $\text{r}_{\text{per}}$ and $\text{r}_{\text{par}}$ values (with $1\sigma$ errorbars) on the mean fraction of surviving orphan galaxies. The red horizontal lines indicate the corresponding values for Andromeda.} 
\label{fraction_means}
\end{figure}

Neither by including all orphan galaxies nor by excluding all of them can we simultaneously fit Andromeda's $\text{r}_{\text{per}}$, $\text{r}_{\text{par}}$ and radial satellite distribution. This might not be too surprising since
its possible that only a fraction of orphan galaxies has survived tidal stripping. In order to further examine the influence of orphan galaxies, we randomly exclude orphan galaxies for each halo to simulate the effect of varying degrees of tidal disruption. Fig. \ref{fraction_means} shows the $\text{r}_{\text{per}}$ and $\text{r}_{\text{par}}$ value averaged over all haloes vs. the fraction of orphan galaxies that is assumed to have survived tidal stripping averaged over all haloes. Both, the mean $\text{r}_{\text{per}}$ and the mean $\text{r}_{\text{par}}$ value, depend almost linear on the mean fraction of orphans with a higher fraction of surviving orphan galaxies leading to a lower mean $\text{r}_{\text{par}}$ and $\text{r}_{\text{par}}$ values. This clearly confirms the conclusion derived above that orphan galaxies are necessary to find thin planes for a significant fraction of 
the haloes. The fraction of haloes with a lower $\text{r}_{\text{per}}$ and an higher $\text{r}_{\text{par}}$ value than Andromeda reaches a maximum if we assume that 69 per cent of all
orphan galaxies survived tidal stripping. In this case  13 per cent of all haloes have a root mean square distance perpendicular to the best fitting plane, $\text{r}_{\text{per}}$,
that is smaller than the corresponding value for the VTPD. In this case the radial distribution is also in good 
agreement with Andromeda, a \mbox{K-S} test yields on average a probability of $\sim 36$ per cent that a profile from Millennium~II and the profile of Andromeda's satellite distribution are drawn from the same distribution.

\subsection{Rotation and spatial evolution of satellite planes} 

\begin{figure}
\begin{center}
\includegraphics[width=95mm]{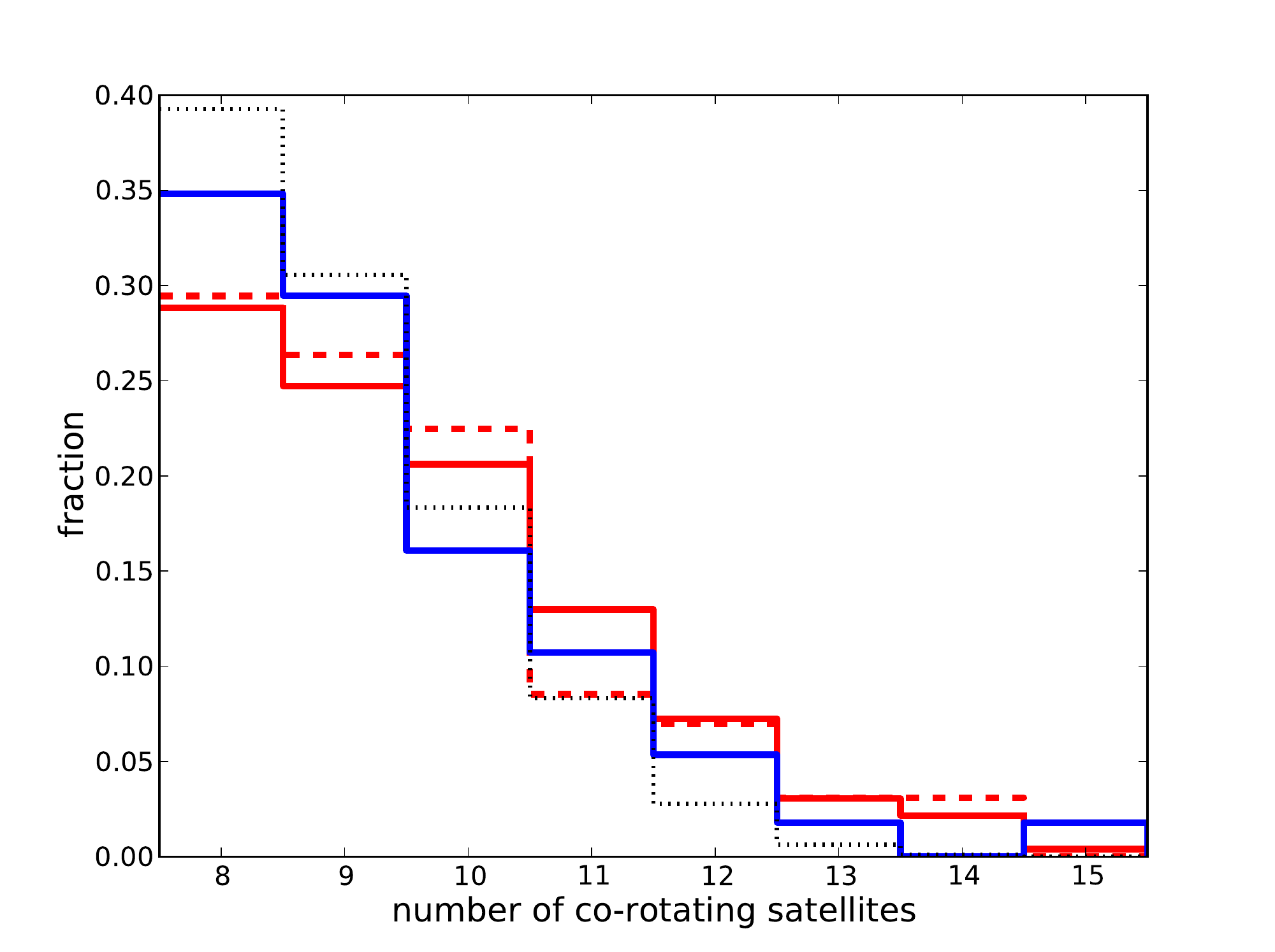}
\end{center}
\caption{Distribution of haloes over the number of co-rotating satellites. The dotted, black line shows the distribution if orbital orientations are randomly distributed (9.1 co-rotating satellites on average). The red, solid line marks the distribution for the Millennium~II haloes if all satellites are considered (9.6 co-rotating satellites on average). The red, dashed line marks the distribution if only haloes with exactly 27 satellites are considered (9.6 co-rotating satellites on average). The blue, solid line marks the distribution for the Millennium~II haloes if orphan galaxies are excluded (9.4 co-rotating satellites on average).} 
\label{rot_analysis}
\end{figure}

In order to understand where the relatively high fraction of haloes with a large number of co-rotating satellite galaxies come from, we finally examine whether the Millennium~II haloes tend to have a larger number of co-rotating satellites compared to a distribution of galaxies with a random orbital orientation. Figure \ref{rot_analysis} shows the fraction of haloes as a function of the number of co-rotating satellites for the Millennium~II haloes - with and without orphan galaxies - and for a distribution with random orbital orientations. Including orphan galaxies, all Millennium~II haloes have on average 9.6 co-rotating satellites. This is slightly higher than the number expected for a purely random distribution (9.1 co-rotating satellites on average), showing that in Millennium~II, dark matter haloes with a mass similar to the virial mass of Andromeda are mildly rotating. If we do not consider orphan galaxies in our analysis, we find that the haloes have on average 9.4 co-rotating satellites. This shows that orphan galaxies tend to move a bit more coherently than non-orphan galaxies, and also that even non-orphan galaxies move more coherently than one would expect if the sense of rotation is random.

\begin{figure*}
\begin{center}
\includegraphics[width=170mm]{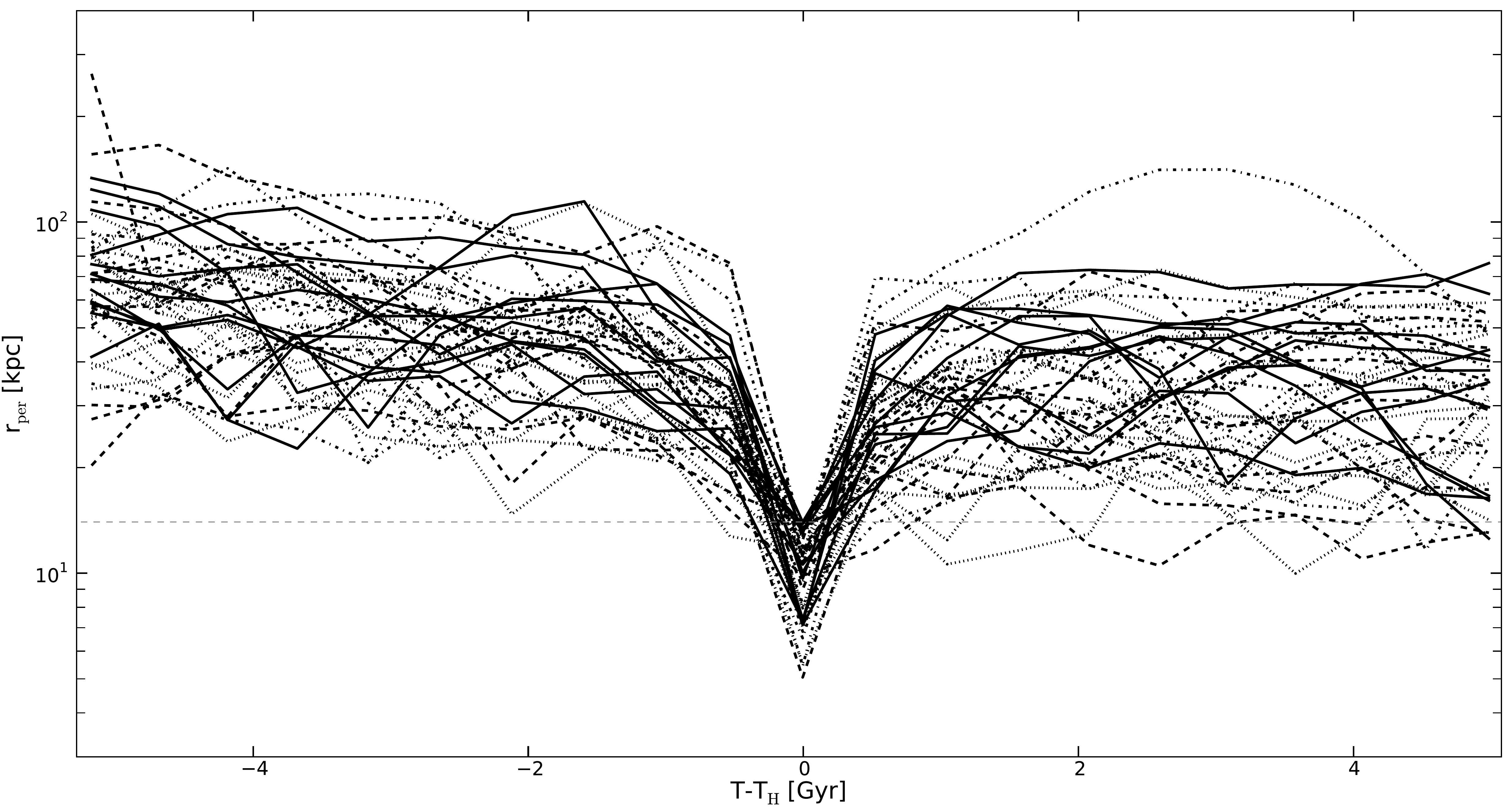}
\end{center}
\caption{Time evolution of the rms distance $\text{r}_{\text{per}}$ of the 15 galaxies which lie closest to the best fitting plane at time $T=T_H$ for all haloes that have $\text{r}_{\text{per}}<14$ kpc at $T=T_H$ and 13 or more co-rotating satellites. The horizontal dashed line marks $\text{r}_{\text{per}}=14$. $\text{r}_{\text{per}}$ drops below 14 kpc and increases again within $\pm$0.5 Gyr. Outside these time intervals it is much larger.} 
\label{form_snap57}
\end{figure*}

We finally test if the satellite planes found in the Millennium~II simulation are real structures, made up of satellite galaxies with tightly aligned orbital planes. To this end,
we investigate the formation and further evolution of the planes for the case that orphan galaxies are included. Fig. \ref{form_snap57} depicts the formation and further evolution of thin planes ($\text{r}_{\text{per}}<$ 14 kpc) with 13 or more co-rotating satellites from $z\sim 0.5$ ($5.2$ Gyr before the present time) up to the end of the simulations at $z\sim -0.3$ (corresponding to $T-T_H = 5.0$ Gyr). The root mean square distance $\text{r}_{\text{per}}$ of the satellites to the best fitting plane decreases slightly from an average value of $\text{r}_{\text{per}}\sim 80$ kpc to $\text{r}_{\text{per}} \sim 30$ kpc from $T-T_H \sim -5.0$ Gyr to $T-T_H \sim -0.5$ Gyr. From $T-T_H \sim -0.5$ Gyr to $0$ Gyr, $\text{r}_{\text{per}}$ drops rapidly from 
an average of $\sim 30$ kpc to values below 14 kpc before increasing again to average values of $\sim 30$ kpc $\sim 0.5$ Gyr after the present time. From $\sim 0.5$ Gyr to $\sim 5$ Gyr the average $\text{r}_{\text{per}}$ values are roughly constant. Individual values fluctuate heavily. Despite the heavy fluctuations, the $\text{r}_{\text{per}}$ values are almost always significantly larger than at $z=0$ when the satellite galaxies belonging to the planes were selected. Consequently, the satellite galaxies can not be tightly aligned in an orbital plane. The timescale over which the average $\text{r}_{\text{per}}$ values are small (about 1 Gyr), agrees roughly with the orbital time of the satellites around their host haloes. We therefore conclude that the thin planes which we have identified in the Millennium~II simulations are mainly statistical fluctuations of an underlying more spherical galaxy distribution.

The analysis presented so far has used data derived by re-scaling the Millennium~II simulation to match WMAP7 cosmology. Such a re-scaling is supposed to reproduce the mass power spectrum of the target cosmology to better than 3 per cent on small scales \citep{A10}. In order to be sure that the final results (if all orphan galaxies are included) are not falsified by the scaling process, we examine additionally the unscaled version of the Millennium~II simulation using the same approach as for the scaled version. The results are in good agreement with the results for the scaled version. For example we find that 44 per cent of the haloes have a smaller $\text{r}_{\text{per}}$ value than Andromeda and 3 per cent a smaller $\text{r}_{\text{per}}$ value than Andromeda and 13 or more co-rotating satellites if orphan galaxies are included, in good agreement with the values obtained in sec. 3.1. Hence our results seem robust against the adopted cosmology.

\section{Conclusion and discussion}

We have compared the observed distribution of Andromeda satellite galaxies, and especially the vast thin plane of co-rotating dwarf galaxies (VTPD) found by \citet{I13}, with a large number of galaxy haloes taken from the Millennium~II simulation. The large sample of considered haloes provides high statistical significance. After imposing several selection criteria (mass cuts, age cuts, \mbox{PAndAS} area) to ensure comparability to the results of \citet{I13} we find planes which are thinner than the VTPD for 40 per cent of the Millennium~II haloes if we include orphan galaxies. We also find that this good agreement is partially due to a mismatch of the radial distributions between the Millennium~II haloes and the satellite system of Andromeda, since the Millennium~II satellites tend to be more radially concentrated than Andromeda's satellites. The thickness of the best fitting plane and the radial concentration of the haloes depend almost linear on the fraction of orphan galaxies that are assumed to have survived tidal stripping. If this fraction is lower, thin planes are found less often but the
radial distribution of the Millennium~II satellites is in better agreement with the Andromeda system. We obtain a best match with Andromeda for an assumed survival fraction 
of orphan galaxies of about 60 per cent. To derive more reliable results will require higher resolution simulations that do not have to consider orphan galaxies when comparing with
the satellite system of Andromeda.

\cite{I13} found that 13 out of 15 satellites belonging to the VTPD share the same sense of rotation. We find that Millennium~II haloes with 13 or more co-rotating satellites orbiting in a thin plane are not rare since we find them in about 2 per cent of all investigated haloes. This number
is significantly higher than expected if the sense of rotation would be random. The reason being that a significant fraction of Millennium~II haloes is mildly rotating. This is in agreement with the results obtained by \citet{L09} and \citet{L11} for the distribution of Milky Way dwarf satellite galaxies.

Several theories have been suggested to explain the formation of thin planes of satellite galaxies. \citet{L05} found that the flat distribution of dwarf galaxies around the Milky Way can be explained by the anisotropic accretion of matter along filaments. \citet{Z05} came to a similar conclusion. \citet{D06} and \citet{D09} developed a scenario in which galaxy formation is driven by cold streams along cosmic filaments. \citet{Go13} found that cold mode accretion streams can explain the formation of the VTPD. Similarly \citet{Sa13} found that the satellites belonging to the VTPD could have all been accreted from an intergalactic filament. In contrast, \citet{H13} modelled Andromeda as an ancient, gas-rich major merger and found that the formation of tidal dwarf galaxies in tails arising from galaxy interactions is the more likely explanation for the existence of the VTPD. A similar scenario has been suggested as the explanation for the plane of satellite galaxies around the Milky Way by \citet{M08} and \citet{M09}.

Our results offer a third possible explanation for the VTPD as it might just be a random statistical fluctuation of an underlying more spherical galaxy distribution. This is indicated by the rapid fluctuation of the $\text{r}_{\text{per}}$ values around $z=0$ shown in Fig. \ref{form_snap57}. A possible instability of the VTPD was also discussed by \citet{B13}, who found that a thin satellite disc can persist over cosmological times ($\sim 5$ Gyr) if and only if it lies in the planes perpendicular to the long or short axis of a triaxial halo, or in the equatorial or polar planes of an oblate or prolate halo. A measurement of the proper motions of Andromeda's satellites would allow to distinguish between the different scenarios but may be very difficult to do.

We have regarded Andromeda's satellite system as an isolated system. The closeness of the Milky Way could indicate a common formation history \citep{H13}. \citet{S13} and \citet{P13} investigated the distribution of all dwarf galaxies in the Local Group. \citet{S13} analysed the formation of four planes containing most of the Local Group dwarfs and found that the key to the formation of the planar structures in the Local Group is the evacuation of the Local Void and consequent build-up of the Local Sheet, a wall of this void. \citet{P13} found that the satellites that do not belong to either Andromeda or the Milky Way can be fitted by two planes and proposed that major galaxy interactions are the key for the understanding of the formation of the four identified planes of Local Group satellite galaxies. Despite this ongoing debate, we conclude that as far as we regard Andromeda's satellite system as an isolated system, the existence of a VTPD is not in conflict with the standard cosmological framework.

\section*{Acknowledgments}

We thank Joel Pfeffer and an anonymous referee for useful suggestions and comments. The Millennium-II Simulation databases used in this paper and the web application providing online access to them were constructed as part of the activities of the German Astrophysical Virtual Observatory (GAVO). H. Baumgardt is supported by the Australian Research Council through Future Fellowship grant FT0991052.

\label{lastpage}

\end{document}